\documentclass[preprint,floats,aps,epsfig,nofootinbib,amssymb]{revtex4-1}
\usepackage{mathrsfs}

\usepackage{slashed}
\usepackage{graphicx,color}
\usepackage{subfigure}
\usepackage{dcolumn}
\usepackage{bm}
\usepackage{subfig}
\usepackage{graphicx}
\usepackage{amssymb}
\usepackage{stackrel}

\begin{document}

\title{Direct Detections of Dark Matter  in the Presence of  Non-standard Neutrino Interactions}

\author{Wei Chao}
\email{chaowei@bnu.edu.cn}
\author{Jian-Guo Jiang}
\email{jgjiang@mail.bnu.edu.cn}
\author{Xuan Wang}
\email{xuanwang@mail.bnu.edu.cn}
\author{Xing-Yu Zhang}
\email{zhangxingyu@mail.bnu.edu.cn}
\affiliation{Center for Advanced Quantum Studies, Department of Physics, Beijing Normal University, Beijing, 100875, China}
\vspace{3cm}

\begin{abstract}

In this paper we investigate impacts of non-standard neutrino interactions (NSIs) to the  limitations on the discovery potential of dark matter in direct detection experiments. New neutrino floors are derived taking into account current upper bounds on the effective couplings of various NSIs.  Our study shows that the neutrino floors of the standard model neutral current interactions can be significantly changed in the presence of vector-current NSI and scalar-current NSI,  and the neutrino floors can be raised up to about ${\cal O}(20\%)$ in the presence of pseudo-scalar-current NSI,  and there are almost no impacts to the neutrino floors from the axial-vector NSI and the tensor NSI.  We suggest combining the dark matter direct detection experiments with the coherent elastic neutrino nucleus scattering experiments to hunt for new physics behind the signal of nuclear recoil  in the future.

\end{abstract}

\maketitle
\section{Introduction}

Cosmological observations have confirmed the existence of dark matter (DM), which points to the new physics beyond the standard model (SM).  Weakly interacting massive particles (WIMP) have been taken as the most attractive DM  candidate, as it can naturally address the observed relic abundance with a weak coupling to the SM particles and an electroweak scale mass. DM direct detection experiments attempt to detect the  recoil energy of nuclei coming from the collisions of nuclei with WIMP in underground laboratories. For the past decades, the detection sensitivity and efficiency of DM direct detection experiments have been greatly improved, but still no signal was observed, which on the other hand puts exclusion limits on the WIMP-nucleus scattering cross section.  It is well-known that the exclusion limits will soon reach the ``neutrino floor"~\cite{Monroe:2007xp,Strigari:2009bq,Billard:2013qya}, the background from  coherent elastic scattering of neutrinos off nuclei. It will be impossible to distinguish the signal of WIMP from that of neutrino using current direct detection techniques  when the signal lies below the neutrino floor.

Several attempts have been made to discriminate the DM signal under the neutrino floor, which include combing data from different targets for WIMP with spin-dependent interactions~\cite{Ruppin:2014bra}, looking for annual modulation~\cite{Davis:2014ama,OHare:2016pjy}, and (or) measuring the recoil momentum~\cite{Grothaus:2014hja,OHare:2015utx}. It has been shown in Ref.~\cite{Gelmini:2018ogy} that it is possible to lift the signal degeneracy associated with the neutrino floor for inelastic scattering.  Actually, we need to understand the neutrino interactions pretty well before making further comparison. Exotic new physics may affect the neutrino floor. It has been shown in Ref.~\cite{Boehm:2018sux} that the neutrino floor can be lifted by several orders of magnitude for DM mass below 10 GeV in the light scalar mediator case,  and a factor of two in light vector mediator case. In Refs.~\cite{Dutta:2017nht,AristizabalSierra:2017joc,Gonzalez-Garcia:2018dep}, authors have studied the effect of non-standard neutrino interactions (NSI) in the DM direct detection experiments.  NSI can enhance or deplete the neutrino-nucleus event rate and thus the neutrino floor can be lifted or submerged. 

In this paper we revisit impacts of NSIs to the  limitations on the discovery potential of dark matter in direct detection experiments.
Recently coherent elastic neutrino-nucleus scattering ($\text{CE}\nu \text{NS}$), predicted by the SM, was observed by the  COHERENT experiment~\cite{Akimov:2017ade}. $\text{CE}\nu \text{NS}$ allows us to study constraints on effective couplings of NSIs. After having considered all updated upper limits,  we evaluated the new neutrino floor induced by the NSIs. Our results show that the neutrino floors of the standard model neutral current interactions can be significantly changed in the presence of vector-current NSI and scalar-current NSI,  and the neutrino floors can be raised up to about ${\cal O}(20\%)$ in the presence of pseudo-scalar-current NSI,  and there are almost no impacts to the neutrino floors from the axial-vector NSI and the tensor NSI.  That is to say, a signal above our new neutrino floors will be definitely that of  DM, while the new physics behind a signal lying between the new and the SM neutrino floors will be blurred and indistinct, in which case one needs combine DM direct detections experiments with $\text{CE}\nu \text{NS}$ experiments to make further identification. For a signal lying below the SM neutrino floors, we need to develop new direct detection methods.

The remaining of the paper is organized as follows: In section II we present the exotic neutrino interactions and cross section of $\text{CE}\nu \text{NS}$ process. Section III is focused on constraints on the effective couplings of NSIs. In section IV we present impacts of these new interactions to the neutrino floor. The last part is concluding remarks. Nuclear response function are listed in the appendix A. 

\section{Non-standard neutrino interactions}

In the SM, $\text{CE}\nu \text{NS}$ is mediated by the Z-boson at the tree level. The solar, atmosphere, accelerator and reactor neutrino oscillation experiments have confirmed that neutrinos are massive and lepton flavors are mixed, which point to new physics beyond the SM.  As a result, neutrinos could interact with SM particles in the presence of new mediators (new gauge bosons or new scalar fields ).
Effective operators induced by these new mediators are called NSI, which is first addressed by  L. Wolfenstein in the consideration of neutrino oscillation in matter~\cite{Wolfenstein:1977ue}. In this section we address all exotic neutrino interactions beyond the original NSI which  is vector-current interactions between neutrinos  and quarks.  It is well-known that there are 16 independent Dirac field bilinears, which can be decomposed into scalar, vector, pseudo-scalar, axial-vector and tensor currents. The most general dimension-6 operators describing effective neutrino-quark interactions can thus be written as
\begin{eqnarray}
{G_F\over \sqrt{2}} \sum_i^{} \bar \nu_\alpha^{} \Gamma^i P_L^{} \nu_\beta^{} \bar q_f^{} \Gamma_i^{} \zeta_i^{}q_f^{}  \label{effective}
\end{eqnarray}  
where $G_F$ is the Fermi constant, $\alpha, \beta$ are flavors of neutrinos, $f(\equiv u,d,s)$ are  flavors of quarks, $\zeta_i$ are dimensionless couplings, and 
\begin{eqnarray}
\Gamma_i = \left\{ \matrix{1,&\gamma^5, &\gamma^\mu, &\gamma^\mu \gamma^5,&\sigma^{\mu\nu}}\right\} \; .
\end{eqnarray}
The  Lagrangian given in Eq.~(\ref{effective}) may come from integrating out new or SM neutral bosons in effective field theory approach.  It would be more accurate if one performs calculation in an ultraviolet (UV) completion model. However effective operator is good enough in investigating low energy neutrino-nucleus scattering and provides a model independent prediction. It should be mentioned that there are higher dimensional (dimension $7$ or $8$) effective operators~\cite{CidVidal:2018eel}, whose constraints as well as effects in DM direct detections  will be presented in a future project. The Wilson coefficients relevant to the standard effective neutrino-quark interactions can be derived by integrating out  $Z$ boson
\begin{eqnarray}
\zeta_{V, ~u(d,s)}^{} =\mp\left( 1-{8(4)\over 3} s_W^2 \right) \delta_{\alpha\beta}^{} \; , \hspace{2 cm} \zeta_{A,~u(d,s)}^{} =\pm  \delta_{\alpha\beta}^{} 
\end{eqnarray}
where $s_W^2=\sin^2 \theta_W^{} \approx 0.238 $, with $\theta_W^{}$ the weak mixing angle.

To calculate the cross section of $\text{CE}\nu \text{NS}$, one needs to match the effective operators given in Eq.~(\ref{effective}) onto effective field theory describing interactions between neutrinos and non-relativistic nucleon, which was done in Refs.~\cite{Fitzpatrick:2012ix,Bishara:2017pfq}. We list in the Table. 1 the relevant form factors,  in which we have ignored the $q^2$ dependence.  $f_{T_q}^N \equiv( \langle N |m_q \bar q q | N\rangle/m_N$) express the light quark contribution to the nucleon mass, $\bar m =(1/m_u+ 1/m_d + 1/m_s)^{-1}$, $\Delta_q^N$ parameterize the quark spin content of the nucleon, $\delta_q^N$ are the difference between the spin of quarks and that of anti-quarks in nucleon, $m_N$ is the mass of nucleon.

\begin{table}[t]
\centering
\begin{tabular}{|l|l|c|}
\hline Quark level &   Nucleon level & Matching conditions  \\ 
\hline
${G_F \over \sqrt{2}} \zeta_{q, S}^{} \bar \nu_\alpha^{}  P_L^{} \nu_\beta^{} \bar q q $& ${G_F \over \sqrt{2}} \zeta_{N,S}^{} \bar \nu_\alpha^{}  P_L^{} \nu_\beta^{} \bar N N $ &  $\zeta_{N, S}=\sum_{q=u,d} \zeta_{q, S} {m_N \over m_q^{} } f_{T_q}^N$ \\
\hline
${G_F \over \sqrt{2}} \zeta_{q, P}^{} \bar \nu_\alpha^{}  P_L^{} \nu_\beta^{} \bar q i\gamma^5 q $& ${G_F \over \sqrt{2}} \zeta_{N,P}^{} \bar \nu_\alpha^{}  P_L^{} \nu_\beta^{} \bar N i \gamma^5 N $ &  $\zeta_{N, P}=\sum_{q=u,d} \zeta_{q, P} {m_N \over m_q^{} }  \left(1- {\bar m \over m_q^{}}\right) \Delta_q^N$ \\
\hline
${G_F \over \sqrt{2}} \zeta_{q, V}^{} \bar \nu_\alpha^{}  \gamma_\mu^{} P_L^{} \nu_\beta^{} \bar q \gamma^\mu q $& ${G_F \over \sqrt{2}} \zeta_{N,V}^{} \bar \nu_\alpha^{} \gamma_\mu^{}  P_L^{} \nu_\beta^{} \bar N \gamma^\mu N $ &  $\zeta_{p, V}=2 \zeta_{u, V} +\zeta_{d, V}$; ~~$\zeta_{n, V}= \zeta_{u, V} +2 \zeta_{d, V}$ \\
\hline
${G_F \over \sqrt{2}} \zeta_{q, A}^{} \bar \nu_\alpha^{}  \gamma_\mu^{} P_L^{} \nu_\beta^{} \bar q \gamma^\mu \gamma^5  q $& ${G_F \over \sqrt{2}} \zeta_{N,A}^{} \bar \nu_\alpha^{} \gamma_\mu^{}  P_L^{} \nu_\beta^{} \bar N \gamma^\mu \gamma^5 N $ &  $\zeta_{N, A}^{} = \sum_q \zeta_{q, A} \Delta_q^N$ \\
\hline
${G_F \over \sqrt{2}} \zeta_{q, T}^{} \bar \nu_\alpha^{}  \sigma_{\mu\nu}^{}P_L^{} \nu_\beta^{} \bar q \sigma^{\mu\nu}  q $& ${G_F \over \sqrt{2}} \zeta_{N,T}^{} \bar \nu_\alpha^{} \sigma_{\mu\nu} P_L^{} \nu_\beta^{} \bar N \sigma^{\mu\nu} N $ &  $\zeta_{N, T}^{} = \sum_q \zeta_{q, T} \delta_q^N$ \\
\hline
\end{tabular}
\caption{ Effective operators from the quark level to the nucleon level and the nucleon form factors.   }\label{nucleon}
\end{table}

Before preceding to the calculation of $\text{CE}\nu \text{NS}$ cross section, one needs to evaluate  effective couplings between neutrinos and  the proton or the neutron. We assume neutrinos couple to u-quark and d-quark universally in NSI, i.e. $\zeta_{u,i}^{} =\zeta_{d,i}^{} =\zeta_i$, since the neutral currents are usually blinded to the isospin. As a result, $\zeta_{p, V}^{} =\zeta_{n, V}^{} =3 \zeta_V^{}  $, $\zeta_{p, A}^{} =\zeta_{n, A}^{}=0.41\zeta_A^{}$ since $\Delta_u^p=\Delta_d^n=0.84$ and $\Delta_d^p=\Delta_u^n=-0.43$~\cite{Gondolo:2004sc}, $\zeta_{p, T}^{} =\zeta_{n, T}^{} =0.61\zeta_T^{}$ as $\delta_u^p=\delta_d^n=0.84$ and $\delta_d^p=\delta_u^n=-0.23$~\cite{Belanger:2013oya}, $\zeta_{p, S} \approx \zeta_{n, S} =16.3\zeta_{ S}$ by using inputs of $f_{T_q}^{p,n}$ provided in~Ref.~\cite{Gondolo:2004sc}, $\zeta_{p,P}^{} \approx 59 \zeta_P$ and $\zeta_{n,P}\approx 55\zeta_P^{}$. 

The differential cross section of $\text{CE}\nu \text{NS}$ in the present of NSIs is calculated in Refs.~\cite{Lindner:2016wff,Altmannshofer:2018xyo}.  For our case, it can be written as
%\begin{eqnarray}
%{d\sigma\over dE_R}=
%&&\left\{(4 E_\nu^2 -2 m_A^{} E_R^{} )\left| \sum_{q=u,d} F_{V}^{q/N } (q^2 ) \zeta_V^{}  \right |^2 + 2 m_A^{} E_R^{}  \left| \sum_{q=u,d} F_{S }^{ q/N  } (q^2 ) \zeta_S^{}  \right |^2 \right\} W_{M}^{00} (q^2) +  \nonumber \\
%&&\left\{ {E_R^{2} m_A^2  \over  m_N^2} (c_N^{} \zeta_P^{})^2 + {E_R^{} \over 8 m_A^{} } (4 E_\nu^2 -2 m_A E_R^{} ) \left(\sum_q \Delta_q^N \zeta_A^{} \right)^2 + 16 E_\nu^2 \left( \sum_q \delta_q^N \zeta_T^{}  \right)^2 \right\} W^{00}_{\Sigma''} (q^2)+ \nonumber \\ 
%&&\left\{  \left( E_\nu^2 - {1\over 2 } m_A^{} E_R^{}  \right)\zeta_A^{N2} + 4 (4E_\nu^2 -2 m_A^{} E_R^{} )\left( \sum_q \delta_q^N \zeta_T^{}  \right)^2  \right\} W_{\Sigma^\prime}^{00} (q^2)
%\end{eqnarray}
\begin{eqnarray}
{d\sigma_\nu\over dE_R}&=& {2G_F^2 m_A \over (2J_A+1) E_\nu^2} \left\{ \sum_{\alpha\beta=0,1} (4 E_\nu^2 -2 m_A^{} E_R^{} )\zeta_V^\alpha \zeta_V^{\beta *} W^{\alpha\beta}_{M} (q^2)  +   \right. \nonumber \\
&&\sum_{\alpha,\beta=0,1} \left( E_\nu^2 + {1\over 2 } m_A^{} E_R^{}  \right)\zeta_{A}^{\alpha}\zeta_A^{\beta*}W_{\Sigma^\prime}^{\alpha\beta} (q^2)+\sum_{\alpha \beta=0,1} {E_R^{} \over 4 m_A^{} } (2 E_\nu^2 - m_A E_R^{} )\zeta_{A}^\alpha \zeta_A^{\beta*}W^{\alpha \beta}_{\Sigma''} (q^2)   +  \nonumber \\
&&8(2E_\nu^2 - m_A^{} E_R^{} )\zeta_{T}^2   W_{\Sigma^\prime}^{00} (q^2)+ 16 E_\nu^2 \zeta_{T}^2 W^{00}_{\Sigma''} (q^2) +2 m_A^{} E_R^{}  \zeta_{S}^2  W_{M}^{00} (q^2) +\nonumber \\
&& \left. \sum_{\alpha\beta=0,1} {E_R^{2} m_A^2  \over   m_N^2} \zeta_P^\alpha \zeta_P^{\beta *} W^{\alpha\beta}_{\Sigma''} (q^2) \right\} \label{master0}
\end{eqnarray}
where $W^{\alpha\beta}_{M,\Sigma', \Sigma^{''}} (q^2)$ are the nuclear response functions, $E_\nu$ is the initial neutrino energy, $E_R$ is the recoil energy of the nucleus, $J_A^{}$ is the spin of target nuclei, $\zeta_X^{0}={1\over 2}(\zeta_{p, X}^{} + \zeta_{n, X}^{} )$ and $\zeta_{X}^1 ={1\over2}(\zeta_{p, X}^{} -\zeta_{n, X}^{})$, $m_A$ is the mass of target nuclei.  Numerical expressions of $W^{\alpha\beta}_{M,\Sigma', \Sigma^{''}} (q^2)$  are given in the appendix A, which are taken from the public code ``\textbf{dmformfactor}" in Ref.~\cite{Anand:2013yka}. 

\section{Constraints}

Before proceeding to the study of neutrino floor, we summarize in this section constraints, on the NSI Wilson coefficients, arising from neutrino oscillations, $\text{CE}\nu \text{NS}$ and deep inelastic scattering (DIS).  According to global fits to oscillation data, one has ~\cite{Esteban:2018ppq,Altmannshofer:2018xyo}
\begin{eqnarray}
&& \zeta_{u,V}^{ee}\in(-0.080,~0.618) \; , \hspace{1cm}\zeta_{d, V}^{ee} \in(-0.012,~0.361) \; ,  \nonumber \\
 && \zeta_{u,V}^{\mu \mu}\in(-0.111,~0.402)\; ,\hspace{1cm}\zeta_{d, V}^{\mu\mu} \in(-0.103,~0.361)\; ,
\end{eqnarray}
at the 95\% C.L.. Constraints of $\text{CE}\nu \text{NS}$ and DIS are separately given by the COHERENT~\cite{Akimov:2017ade,Akimov:2018ghi} and CHARM~\cite{Qian:2018wid} collaborations. Since these constraints were already studied in references, we will not repeat the investigation here.  We list in the Table.~\ref{limits}  the most stringent current or predicted constraints on $\zeta_{q, X}$, which are derived by translating results of Table. II and III in Ref.~\cite{Altmannshofer:2018xyo} into the upper bounds of $\zeta_{q, X}$ in our case.

\begin{table}[t]
\centering
\begin{tabular}{c|c||c|c||c|c||c|c}
\hline
\hline Couplings &   Constraints  & Couplings &   Constraints&Couplings & Constraints&Couplings & Constraints  \\ 
\hline
$\zeta^{eX}_{u, S}$&0.051& $\zeta^{\mu X}_{u,S}$&0.035&$\zeta^{eX}_{u, P}$&4.863& $\zeta^{\mu X}_{u,P}$&0.484\\
$\zeta^{eX}_{d, S}$&0.051& $\zeta^{\mu X}_{d,S}$&0.034&$\zeta^{eX}_{d, P}$&6.256& $\zeta^{\mu X}_{d,P}$&0.686\\
$\zeta^{eX}_{s, S}$&0.866& $\zeta^{\mu X}_{s,S}$&0.579&$\zeta^{eX}_{s, P}$&11.87& $\zeta^{\mu X}_{s,P}$&1.603\\
\hline
$\zeta_{u, T}^{eX}$ &0.632 & $\zeta^{\mu X}_{u, T}$ & 0.064& $\zeta_{u, A}^{eX}$ &0.996 &$\zeta_{u, A}^{\mu X}$ & 0.178\\
$\zeta_{d, T}^{eX}$ &0.866 & $\zeta^{\mu X}_{d, T}$ &0.093 & $\zeta_{d, A}^{eX}$ &0.996 &$\zeta_{d, A}^{\mu X}$ &0.250 \\
$\zeta_{s, T}^{eX}$ &1.680 & $\zeta^{\mu X}_{s, T}$ & 0.215& $\zeta_{s, A}^{eX}$ &2.123 &$\zeta_{s, A}^{\mu X}$ & 0.500\\
\hline
$\zeta^{eX}_{u,V}$& 0.123& $\zeta^{\mu X}_{u, V}$ & 0.084\\
$\zeta^{eX}_{d,V}$& 0.112& $\zeta^{\mu X}_{d, V}$ & 0.072\\
$\zeta^{eX}_{s,V}$& 2.123& $\zeta^{\mu X}_{s, V}$ & 0.566\\
\hline
\hline
\end{tabular}
\caption{ Upper limits on  the effective couplings.   }\label{limits}
\end{table}

\section{Neutrino floor}

DM direct detection experiments, which are designed to search for the nuclear recoil in the scattering of WIMPs off nuclei, probe DM straightforwardly. There are many on the running or designed DM direct detection experiments on the world,  for a review of direct detection experiments and their current status, see~\cite{Undagoitia:2015gya,Liu:2017drf} and references therein for detail.  The WIMP event rate can be written as~\cite{Lewin:1995rx}
\begin{eqnarray}
{d R \over d E_R} = {MT} \times  {\rho_{\rm DM  }\sigma^0_n A^2   \over  2 m_{\rm DM} \mu_n^2}   F^2 (E_R^{} ) \int_{v_{\rm min}}  {f(\vec{v}) \over v } d^3 v \label{mastereq}
\end{eqnarray}
where $M$ is the target mass, $T$ is the exposure time, $\rho_{ \rm DM} =0.3 $ ${\rm GeV/c^2/cm^3}$ being the DM density in the local halo, $\mu_n$ is the nucleon-DM reduced mass, $\sigma_n^0$ is the DM-nucleon cross section, $A$ is the atomic number, $F(E_R^{})$ is the nuclear form factor and we use the Helm form factor~\cite{Helm:1956zz}, $f(\vec{v})$ is assumed to be the Maxwell-Boltzmann distribution function describing the DM velocity distribution in the Earth frame, $v_{\rm min}$ depends on $E_R$: $v_{\rm min} =\sqrt{m_N^{} E_R^{} /2\mu_N^2}$ with $\mu_N$ the DM-nucleus reduced mass.  The velocity integral in eq.~(\ref{mastereq}) can be analytically written as~\cite{Barger:2010gv}
\begin{eqnarray}
\int_{v_{\rm min}}  {f(\vec{v}) \over v } d^3 v  = {1\over 2 v_0^{}  \eta_E^{} } \left[{\rm erf} (\eta_+^{} ) - {\rm erf} (\eta_-^{} ) \right]- {1\over \pi v_0  \eta_E^{} }(\eta_+^{} -\eta_-^{} ) e^{-\eta_{\rm esc}^2}
\end{eqnarray}
where $v_0$ is the speed of the Local Standard of Rest, $\eta_E=v_E/v_0$ with $v_E$ the Earth velocity with respect to the galactic center, $\eta_{\rm esc} =v_{\rm esc} /v_0$ with $v_{\rm esc}$  the escape velocity of DM from our galaxy, $\eta_{\pm} ={\min} (v_{\rm min}/v_0 \pm \eta_E, v_{\rm esc}/v_0)$.  We take $v_0=220~{\rm km/s}$, $v_{\rm esc} =544~{\rm km/s}$
and $\vec{v}_E = \vec{v}_\odot +\vec{v}_\oplus \approx \vec{v}_\odot =232~{\rm km/s}$, where $\vec{v}_\odot$ and $\vec{v}_\oplus$ are the velocity of the sun with respect to the Galaxy as well as the Earth rotational velocity, respectively. 

Although much efforts are made to improve the detection sensitivity and efficiency,  no DM signal was observed in any direct detection experiments. The 90\% CL upper bound on the zero observed counts is 2.3 event~\cite{Goodstein:2012zz}, so we can get the exclusion limit on the direct detection cross section  in the $m_{\rm DM}-\sigma_n^0$ plane for a concrete direct detection experiment with fixed exposure,  by requiring $\int dR/dE_R \varepsilon(E_R^{} ) d E_R^{} <2.3$, where $\varepsilon(E_R)$ is the detector efficiency function and is set to be 1 in our following analysis.

It is well-known that the exclusion of the spin-independent direct detection cross section will soon reach the neutrino floor, below which the spectrum of the recoil energy induced by WIMP-nucleus scattering can not be distinguished from that induced by the $\text{CE}\nu \text{NS}$.  The background is due to solar neutrinos at low recoil energies and atmosphere neutrinos or supernovae neutrinos  at high recoil energies.  Some approaches were proposed on how to extract DM signature from below the neutrino floors. Here we focus on the neutrino floor itself and evaluate the impacts of exotic neutrino interactions to the neutrino floors.  The event rate induced by the $\text{CE}\nu \text{NS}$ can be written as
\begin{eqnarray}
{d R_\nu \over d E_R^{} } = MT \times {1\over m_A^{} } \int_{E_\nu^{\rm min}} {d\phi_\nu \over d E_\nu} {d \sigma_\nu \over d E_R^{} } d E_\nu^{} 
\end{eqnarray} 
where $d\phi_\nu/d E_\nu$ is the flux of neutrinos, $d\sigma_\nu/ d E_R$  is the differential cross section of $\text{CE}\nu \text{NS}$ given in Eq. (\ref{master0}). The relevant flux used in our analysis are taken from Refs.~\cite{Bahcall:2004mq,Battistoni:2005pd}.  The minimum energy of neutrinos, $E_\nu^{\rm min}$, required to induce a nuclear recoil at the energy $E_R$ is $\sqrt{m_A^{} E_R^{} /2}$.

\begin{figure}[t]
\includegraphics[width=0.45\textwidth]{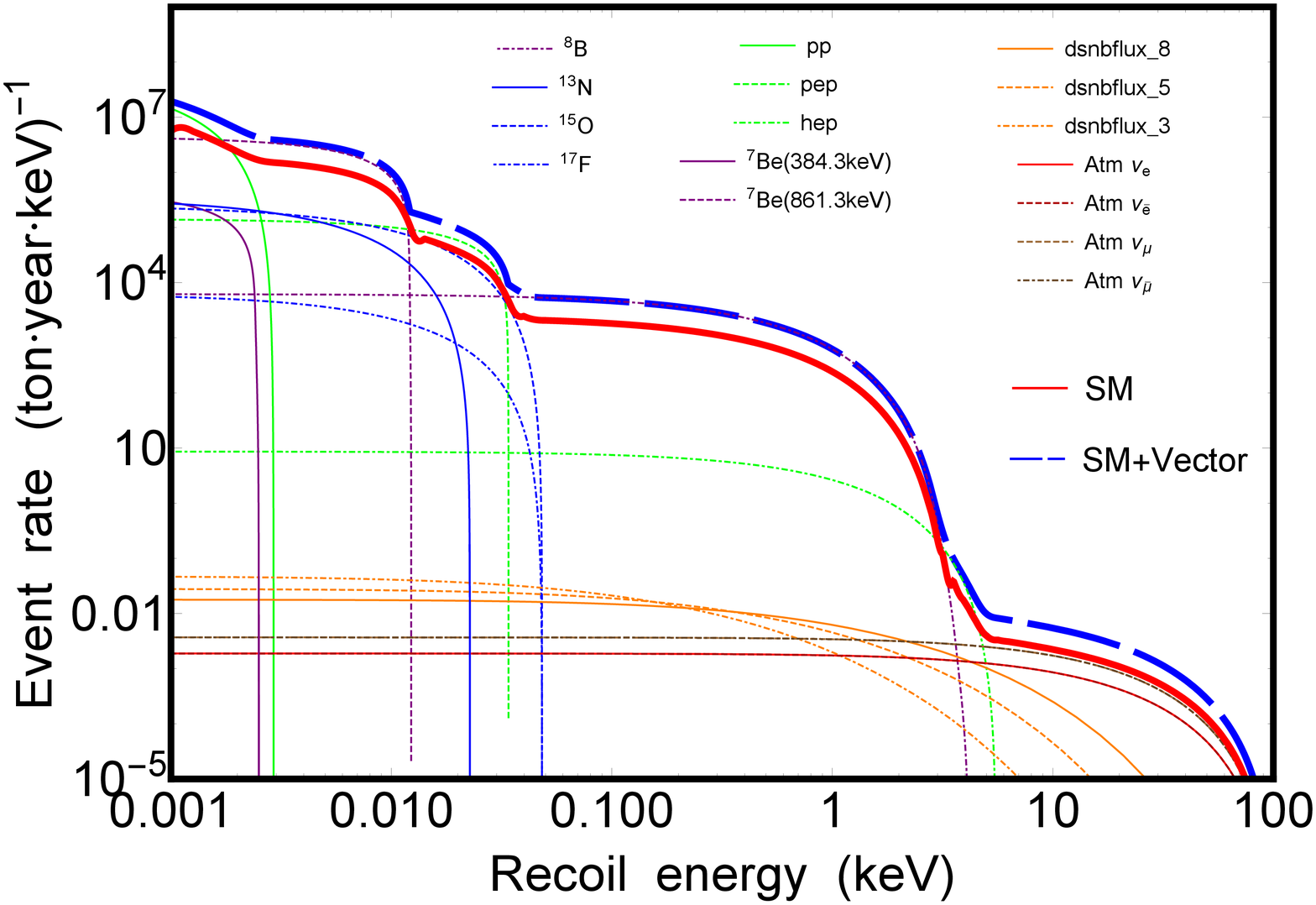}
\hspace{0.5cm}
\includegraphics[width=0.45\textwidth]{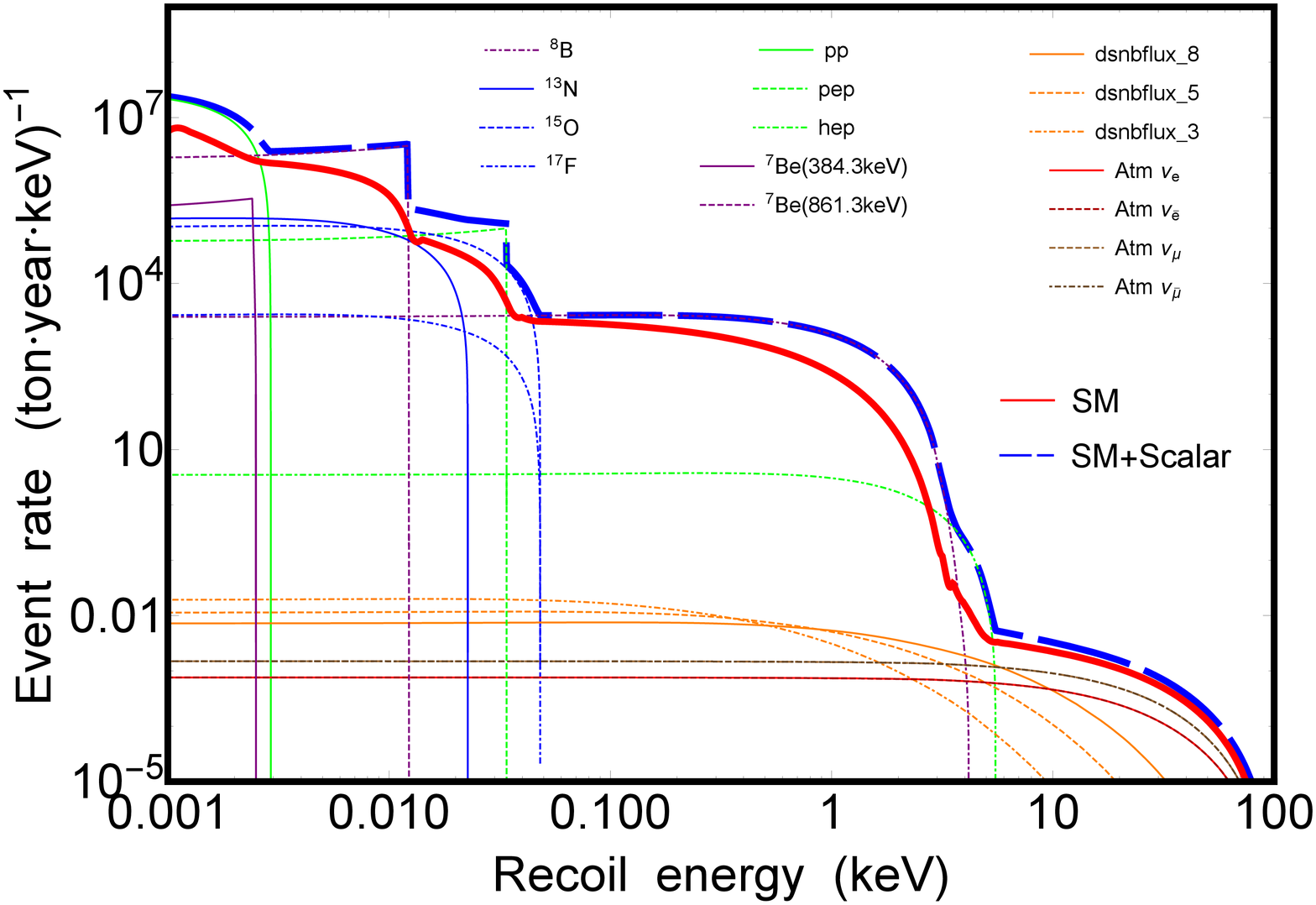}
\includegraphics[width=0.45\textwidth]{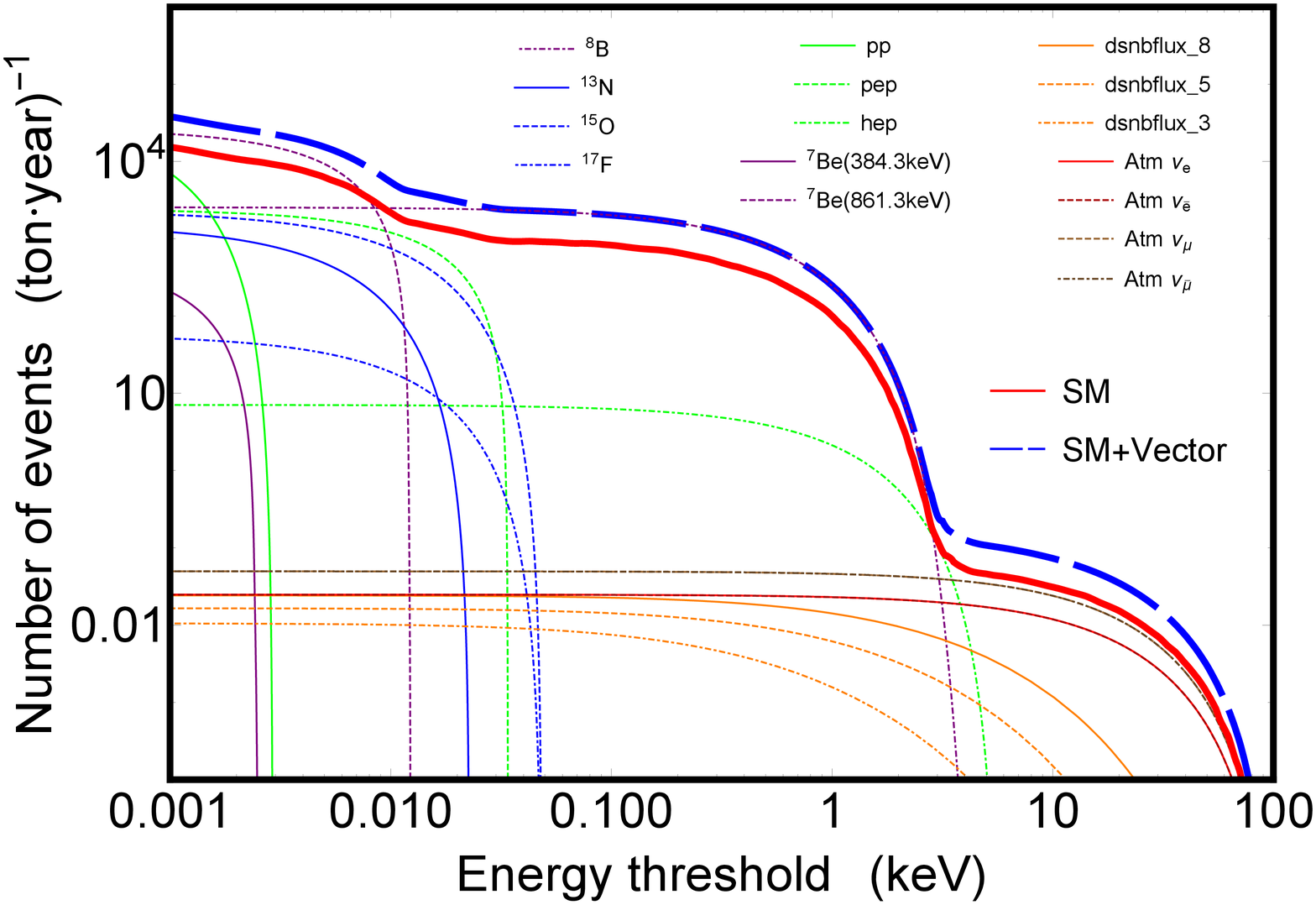}
\hspace{0.5cm}
\includegraphics[width=0.45\textwidth]{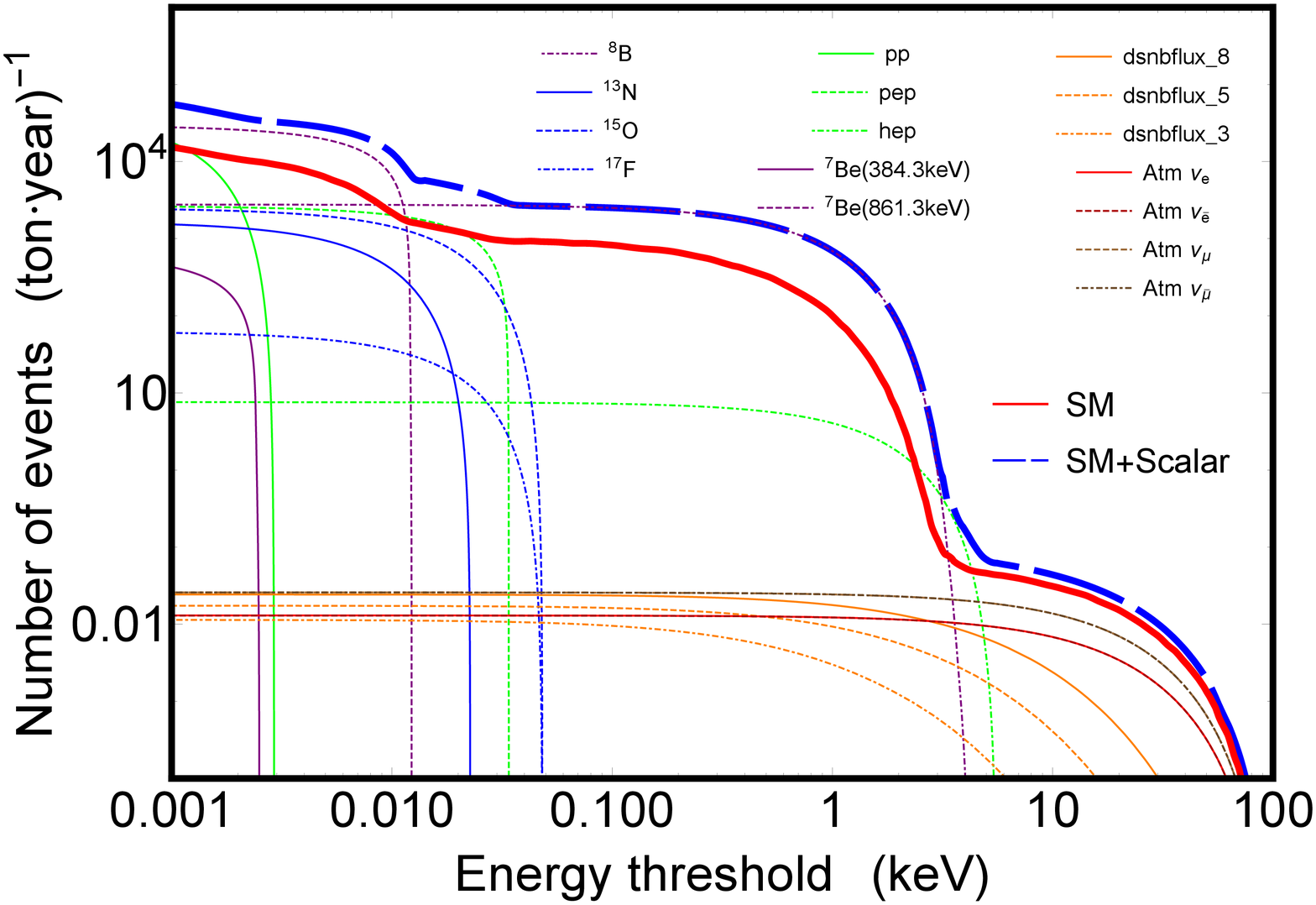}
\includegraphics[width=0.45\textwidth]{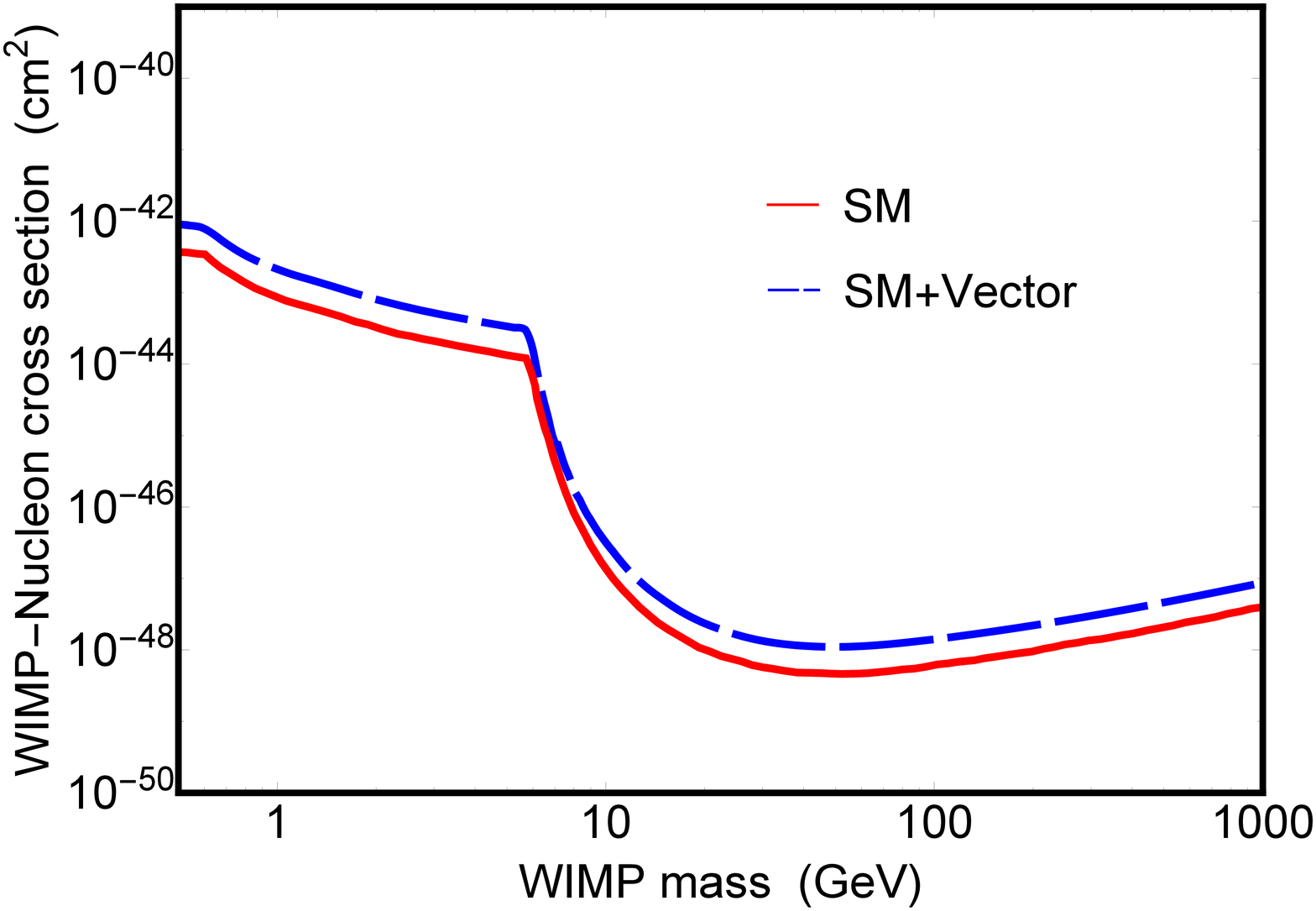}
\hspace{0.5cm}
\includegraphics[width=0.45\textwidth]{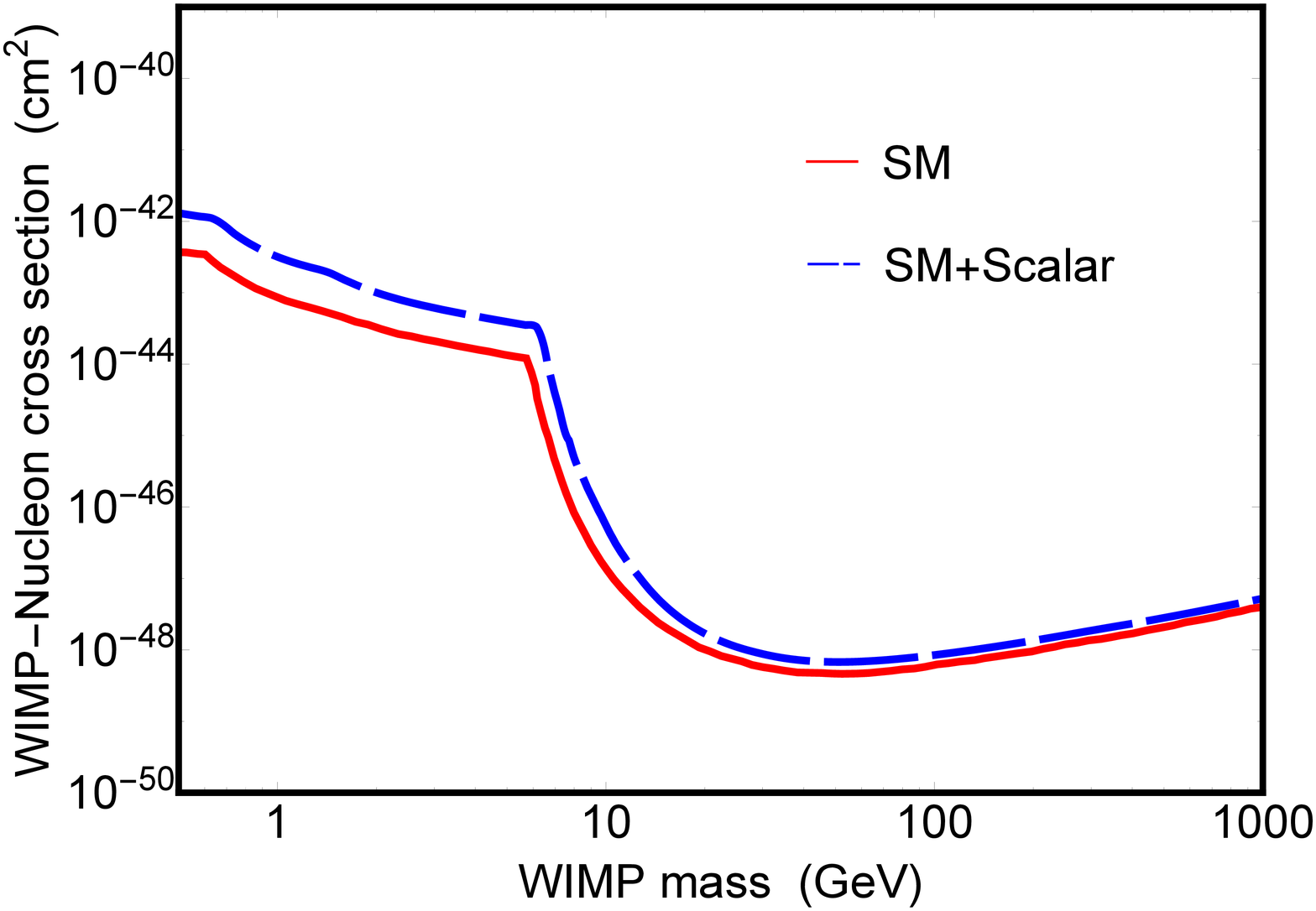}
\caption{First row: event rate as the function of the recoil energy in the Xe131 target in the presence of a single vector current NSI (left-panel) and single scalar current NSI (right-panel). Second row: number of event as the function of energy threshold in the presence of a single vector current NSI (left-panel) and single scalar current NSI (right-panel).  Third row: neutrino floor as the function of dark matter mass  in the presence of a single vector current NSI (left-panel) and single scalar current NSI (right-panel).   } \label{fig:eventrate}
\end{figure}

For a given target, one can construct the neutrino floor in the following way: First, calculating the exposure required to generate $n$ counts of  $\text{CE}\nu \text{NS}$ for a given minimum energy threshold. Second, computing the spin-independent WIMP-nucleon cross section using the following master equation,
\begin{eqnarray}
\sigma_n^0 = {2.3 \over n} \int_{E_R}\left(  {1\over m_N^{} } \int_{E_\nu^{\rm min}} {d\phi_\nu \over d E_\nu} {d \sigma_\nu \over d E_R^{} } \right) \left({\rho_{\rm DM  } A^2   \over  2 m_{\rm DM} \mu_n^2} \int_{E_R^{}}^{E_R^{\rm max}}  F^2 (E_R^{} )  d E_R^{} \int_{v_{\rm min}}  {f(\vec{v}) \over v } d^3 v  \right)^{-1} \; .
\end{eqnarray}
With this equation, one can get the neutrino floor with $n$ neutrino events and estimate the impactions of NSIs to the direct detections of DM.

\section{results}

Taking into account the the upper bound on the effective couplings of NSI, the cross section of $\text{CE}\nu \text{NS}$ can be significantly changed. We first  evaluate impacts of  NSIs to the neutrino event rate.  In  the first row of the Fig.~\ref{fig:eventrate}, we show the event rate as the function of the recoil energy in the Xe131 target in the presence of a single NSI, where plots in the left-panel and right-panel correspond to the vector-current NSI and the scalar-current NSI respectively.  The red solid lines in both plots are the cases of SM neutrino interaction, while the blue dashed lines are cases with additional NSI. Notice that  the shape of the curve with additional vector-current NSI is similar to that of the SM case, while the behavior of of the curve with additional scalar-current NSI is very different from the SM one. This is because the vector-current NSI interferes with the SM contribution and only change the size of the effective coupling, while the scattering cross section from the scalar-current NSI is only coherently enhanced and there is no interference with the SM contribution. In addition, dependences of the cross section on the recoil energy in scalar-current and vector-current  are different, as can be seen from the Eq.~(\ref{master0}). 

We show in the second row of the Fig.~\ref{fig:eventrate} number of $\text{CE}\nu \text{NS}$ events generated within one ton$\cdot$year exposure  in the Xe131 target as the function of  energy threshold for vector-current NSI (left-panel) and scalar-current NSI (right-panel) respectively.  We have set an upper bound (100 keV) on the nuclear recoil energy  when making the plot. Plots may be changed according to this upper limit. Number of events can be significantly enhanced for both low and high energy threshold in the presence of vector-current NSI, and number of events is only enhanced for low energy threshold in presence of  scalar-current NSI.  It is because there is no interference between the SM contribution and that from the scalar-current NSI, thus the relative enhancement will significantly decrease with the increase of the threshold energy. 

\begin{figure}
\includegraphics[width=0.45\textwidth]{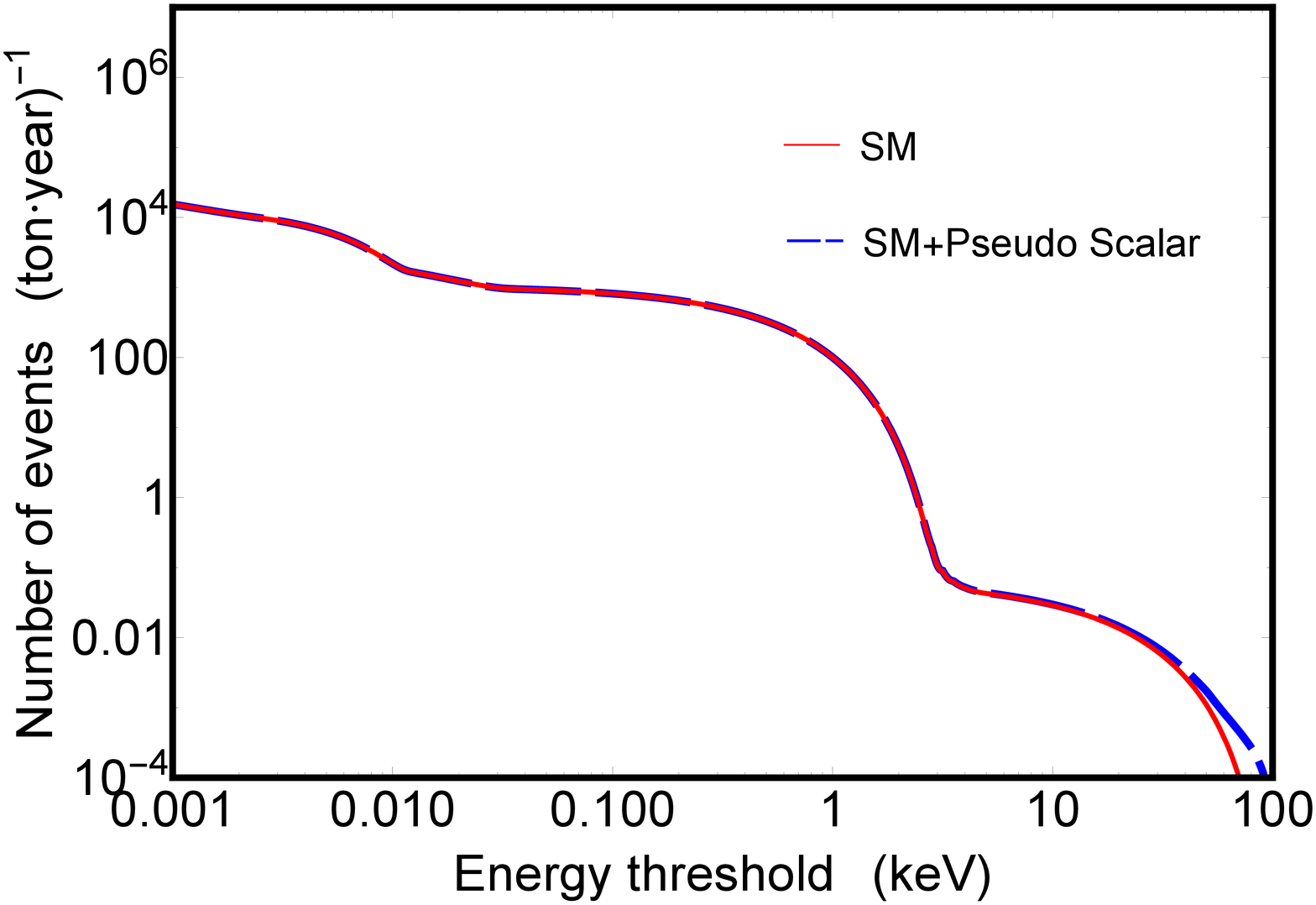}
\hspace{0.5cm}
\includegraphics[width=0.45\textwidth]{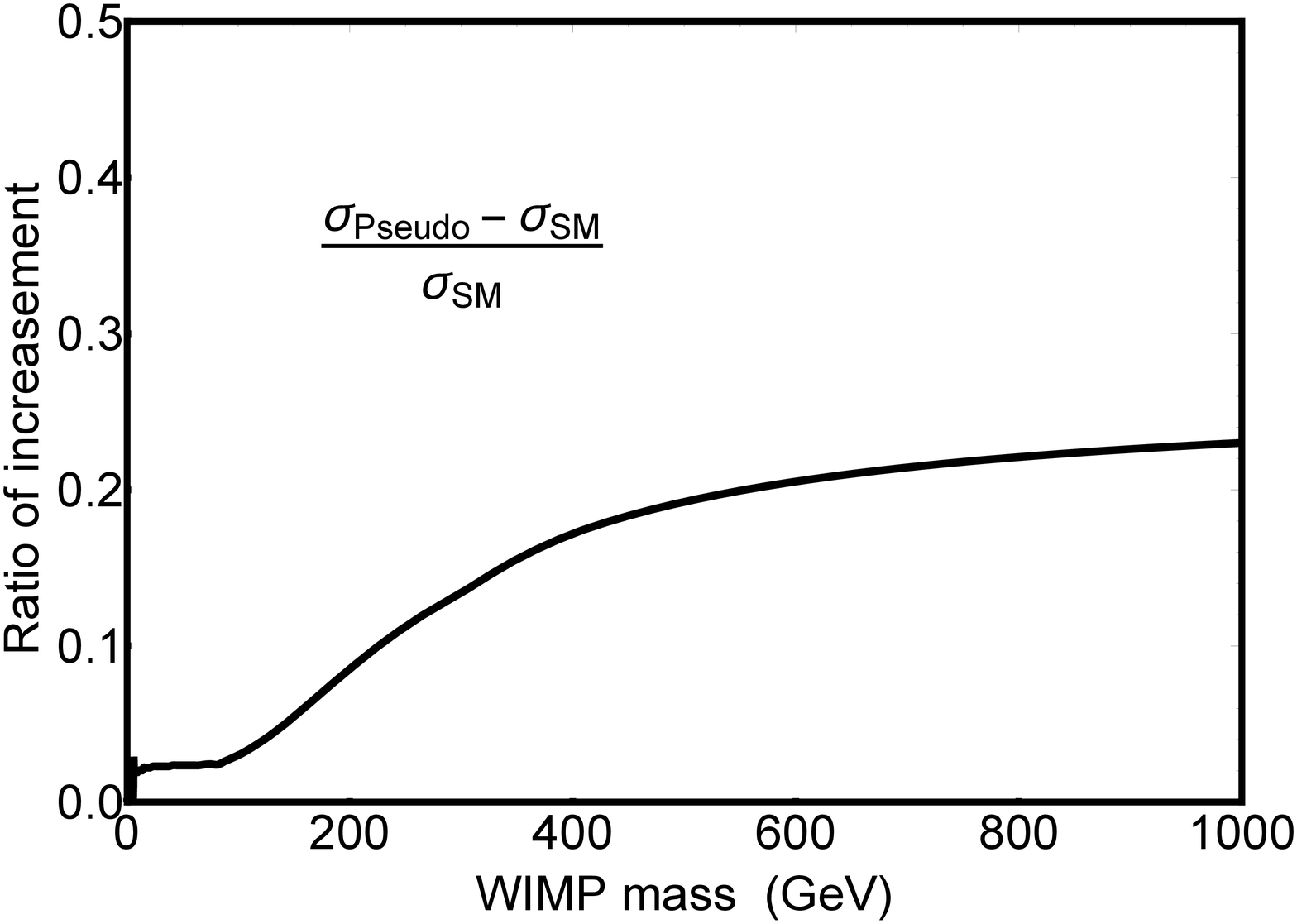}
\caption{Left-panel:  neutrino events as the function of the energy threshold in the Xe-131 detector in the presence of a single pseudo-scalar current NSI; right-panel: enhancement of the neutrino floor with respect to the SM case.  } \label{fig:pseudo}
\end{figure}

We show in the third row of the Fig.~\ref{fig:eventrate} the neutrino floor  in the Xe131 target as the function of dark matter mass for vector-current (left-panel) NSI and scalar-current (right-panel) NSI respectively.  The red solid lines are the SM case and  the blue dashed lines are cases with additional NSIs. Each point in  curves corresponds to the WIMP-nucleon scattering cross section in the exposure where one $\text{CE}\nu \text{NS}$ event is generated that can not be distinguished from the WIMP event.  To make the plot, we first evaluate the exposure such that Xe131 target expects one neutrino event,  with varying energy threshold from $10^{-3}$ keV to $100$ keV. Then we calculate the WIMP-nucleon scattering cross section for each exposure by requiring $\int dR/dE_R \varepsilon(E_R^{}) =2.3$. By taking the smallest cross section for various energy threshold at a fixed dark matter mass,  that corresponds to the best background free sensitivity estimate achievable,  one can draw the curve.  As can be seen, neutrino floors can be significantly raised by the vector-current NSI in both low and high dark matter mass region, and the neutrino floors can be raised by scalar-current NSI only in low dark matter mass region.

We show in the left-panel of the Fig.~\ref{fig:pseudo}  number of neutrino event within 1 ton$\cdot$year exposure in the Xe131 target as the function of energy threshold for pseudo-scalar-current NSI (blue dashed line). As can be seen, the enhancement is tiny for small energy threshold, this is because the contribution of the pseudo-scalar current NSI  to the $\text{CE}\nu \text{NS}$ is suppressed by the  tiny nuclear response functions $W_{\Sigma''}^{\alpha \beta} (q^2)$. For a large recoil energy, its effect become significant because the contribution of the pseudo-scalar-current NSI to the $\text{CE}\nu \text{NS}$ cross section is proportional to $E_R^2$.  We show in the right-panel of the Fig.~\ref{fig:pseudo} a ratio $R$ as the function of dark matter mass, with $R$ defined by 
\begin{eqnarray}
R= {\sigma_{\rm neutrino ~floor}^{\rm SM+NSI} -\sigma_{\rm neutrino ~floor}^{\rm SM} \over \sigma_{\rm neutrino~ floor}^{\rm SM}} \label{ratio}
\end{eqnarray}
where $\sigma_{\rm neutrino ~floor}^{\rm SM+NSI}$ is the neutrino floor with pesudo-scalar-current NSI and $\sigma_{\rm neutrino ~floor}^{\rm SM}$ is the neutrino floor induced by the SM neutral current interactions. It is clear that the enhancement can be of ${\cal O}(20\%)$. 

\begin{figure}
\includegraphics[width=0.47\textwidth]{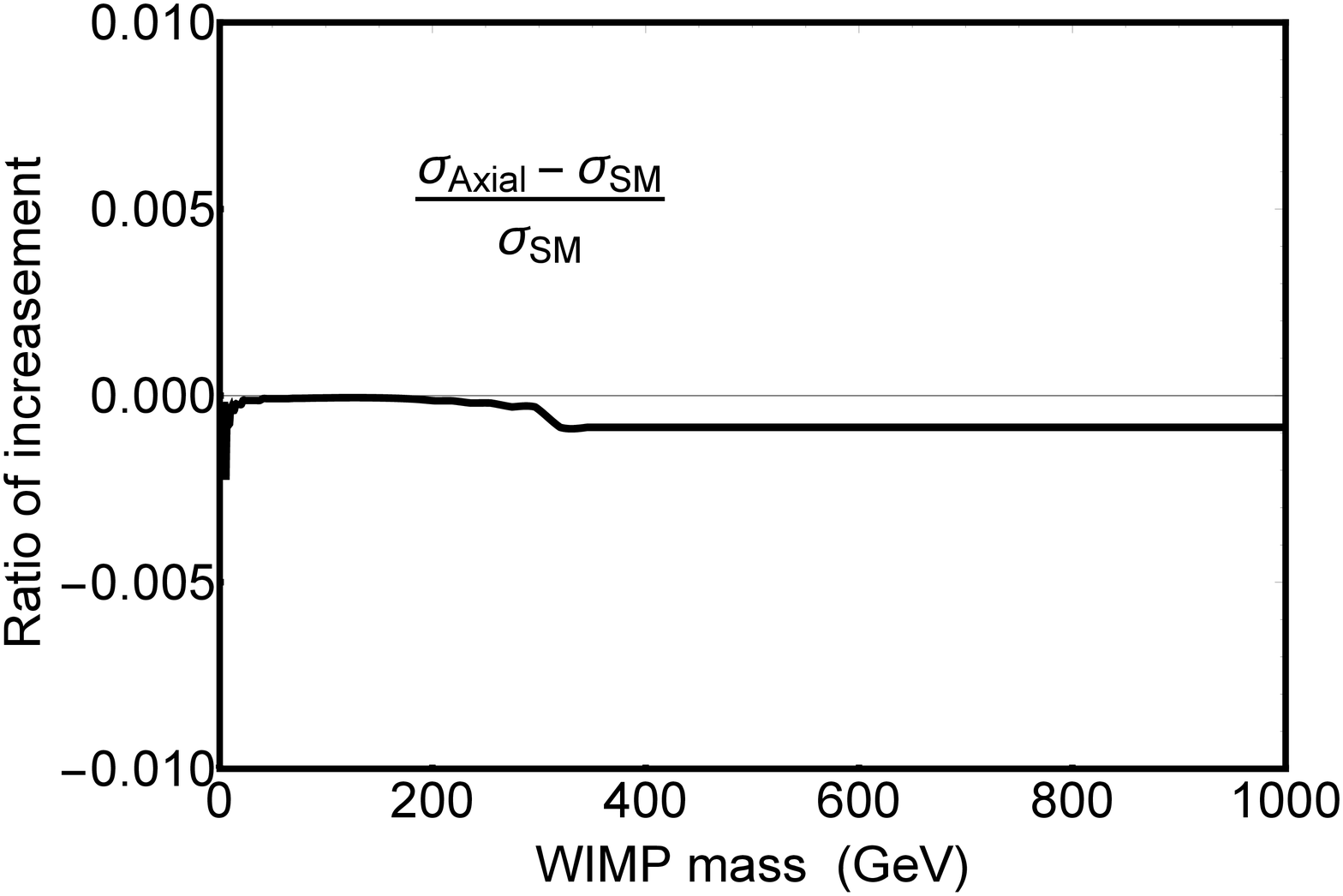}
\hspace{0.5cm}
\includegraphics[width=0.47\textwidth]{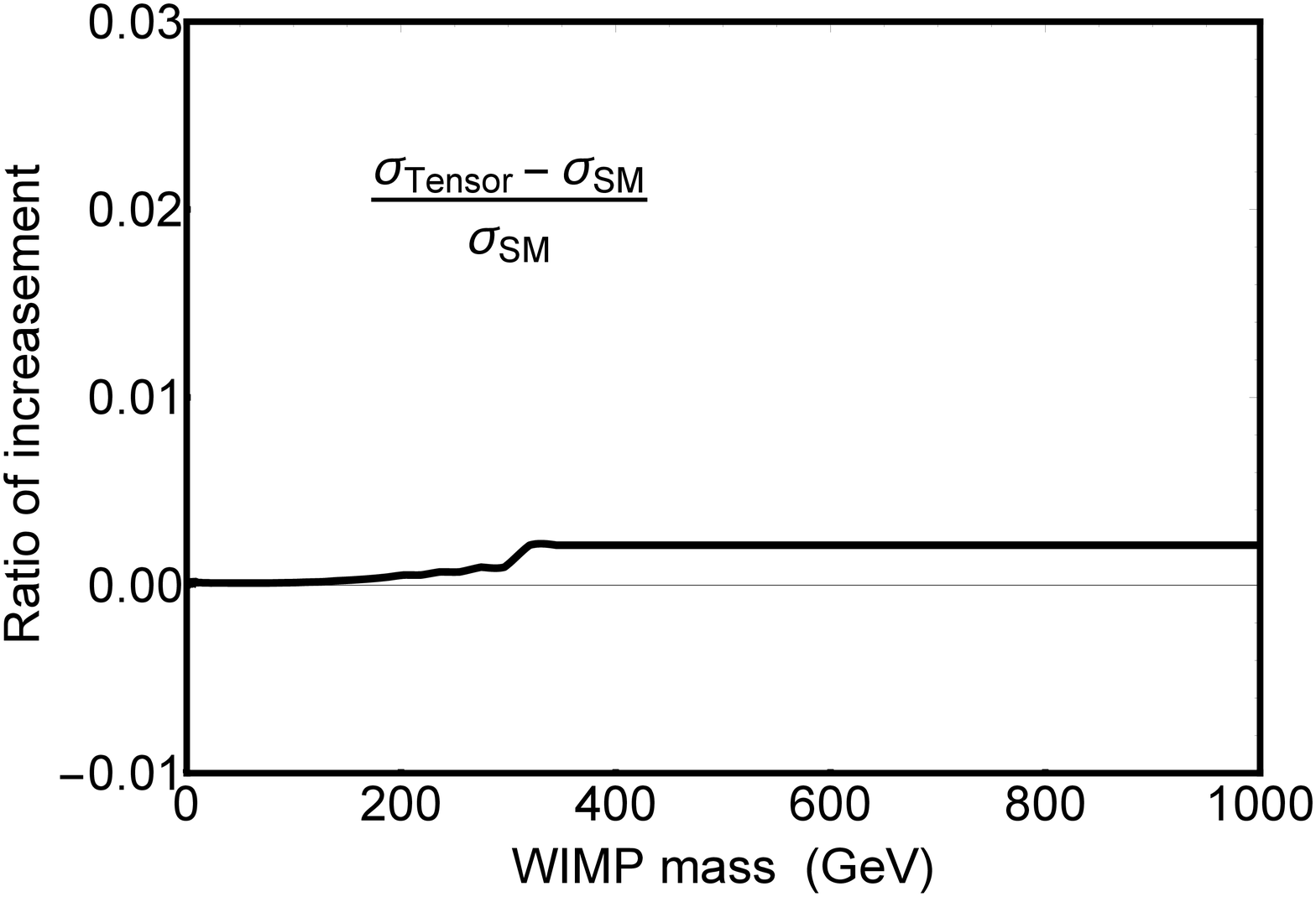}
\caption{The ratio $R$ as the function of dark matter mass for the axial-vector current NSI (left-panel) and tensor current NSI (right-panel).   } \label{fig:tensoraxial}
\end{figure}

We find that contributions of the the axial-vector-current NSI and the tensor-current NSI to the $\text{CE}\nu \text{NS}$ are suppressed by the nuclear response function and the enhancement to the neutrino events can be neglected. As illustrations, we show in the Fig.~\ref{fig:tensoraxial} the ratio $R$,  which is defined in Eq. (\ref{ratio}), as the function of dark matter mass for the axial-vector-current NSI(left-panel) and tensor-current NSI (right-panel). Changes of the neutrino floor due to  these two kinds of NSIs are within ${\cal O} (1\%)$. 

\begin{figure}[t]
\includegraphics[width=0.45\textwidth]{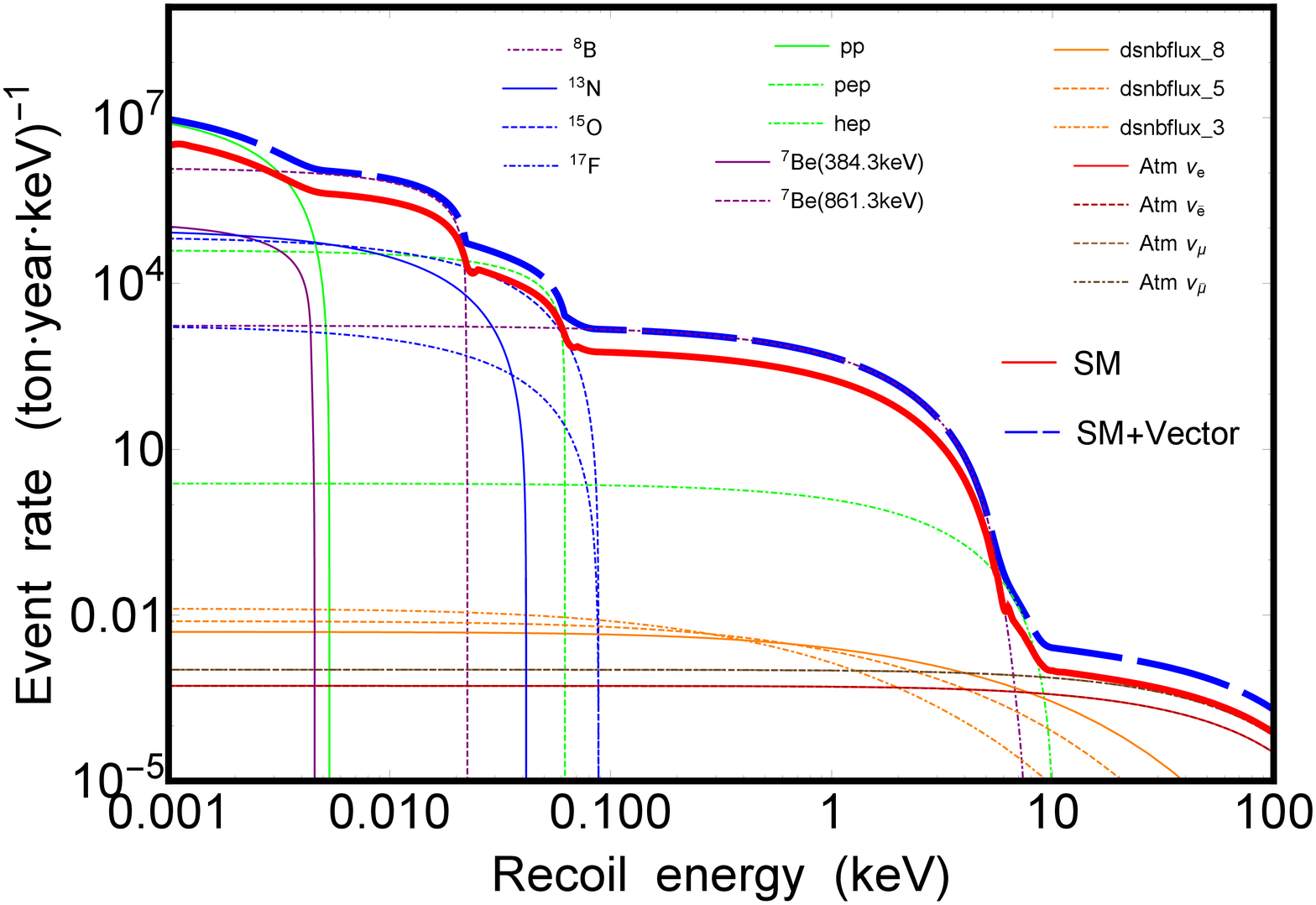}
\hspace{0.5cm}
\includegraphics[width=0.45\textwidth]{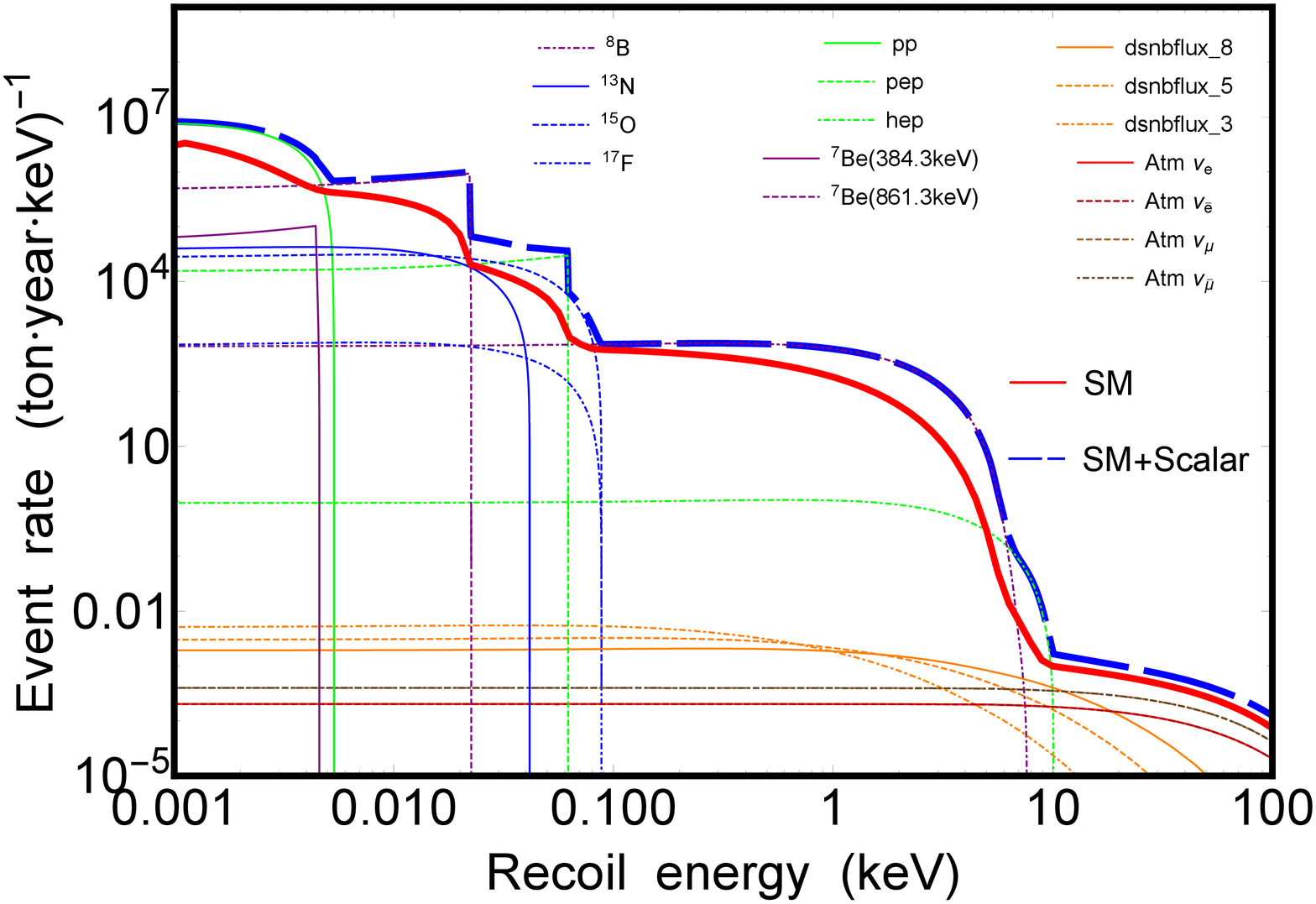}
\includegraphics[width=0.45\textwidth]{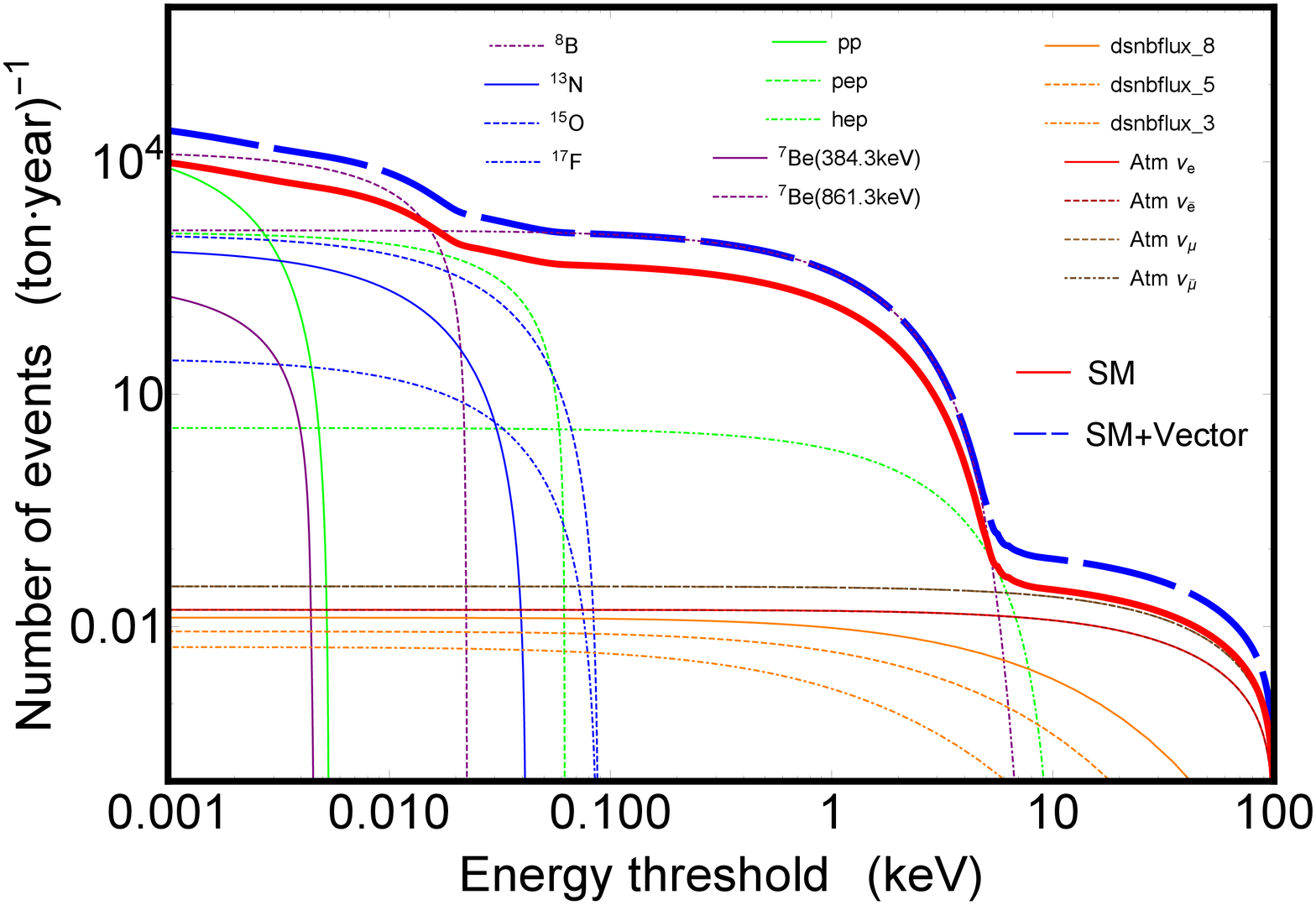}
\hspace{0.5cm}
\includegraphics[width=0.45\textwidth]{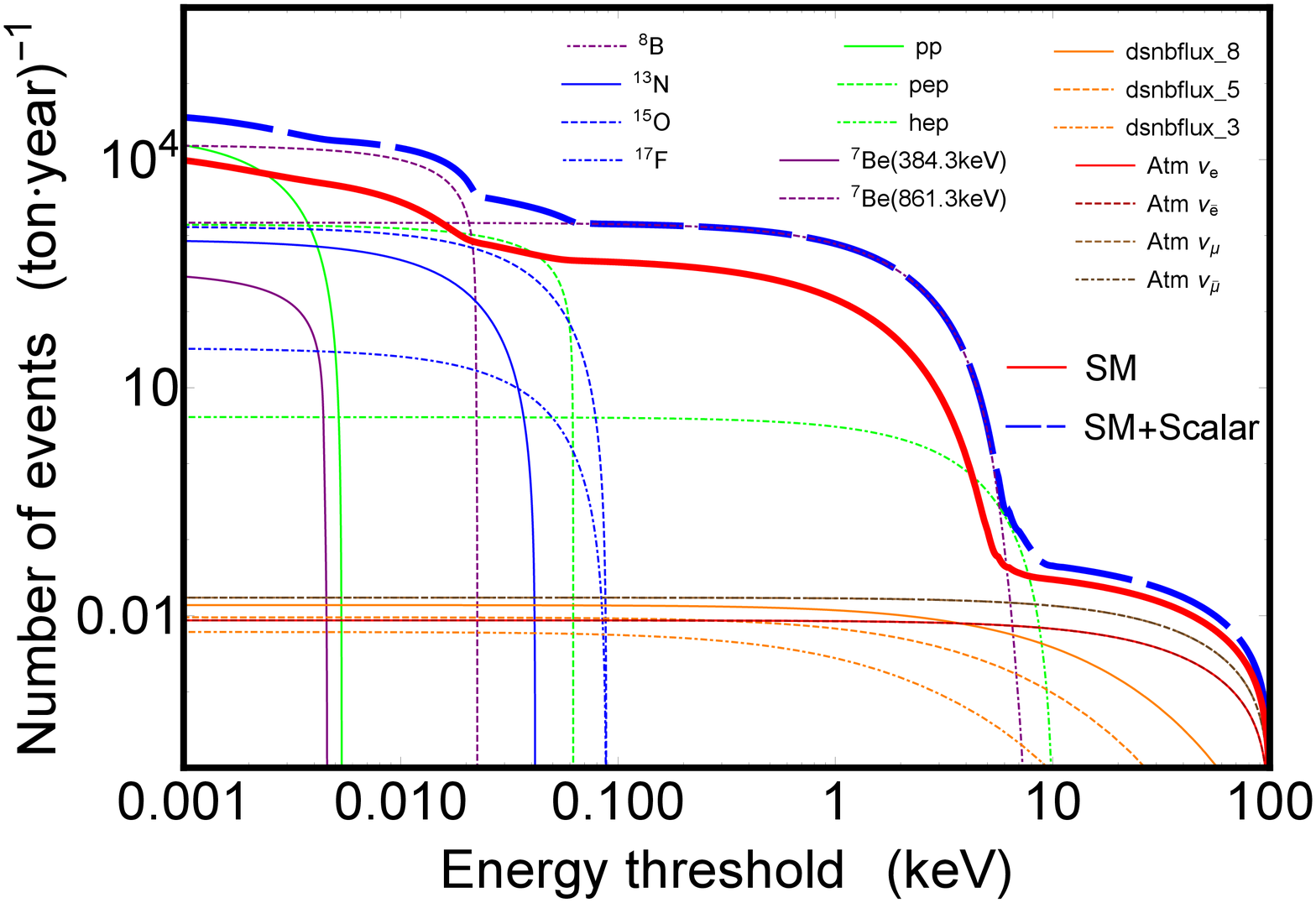}
\includegraphics[width=0.45\textwidth]{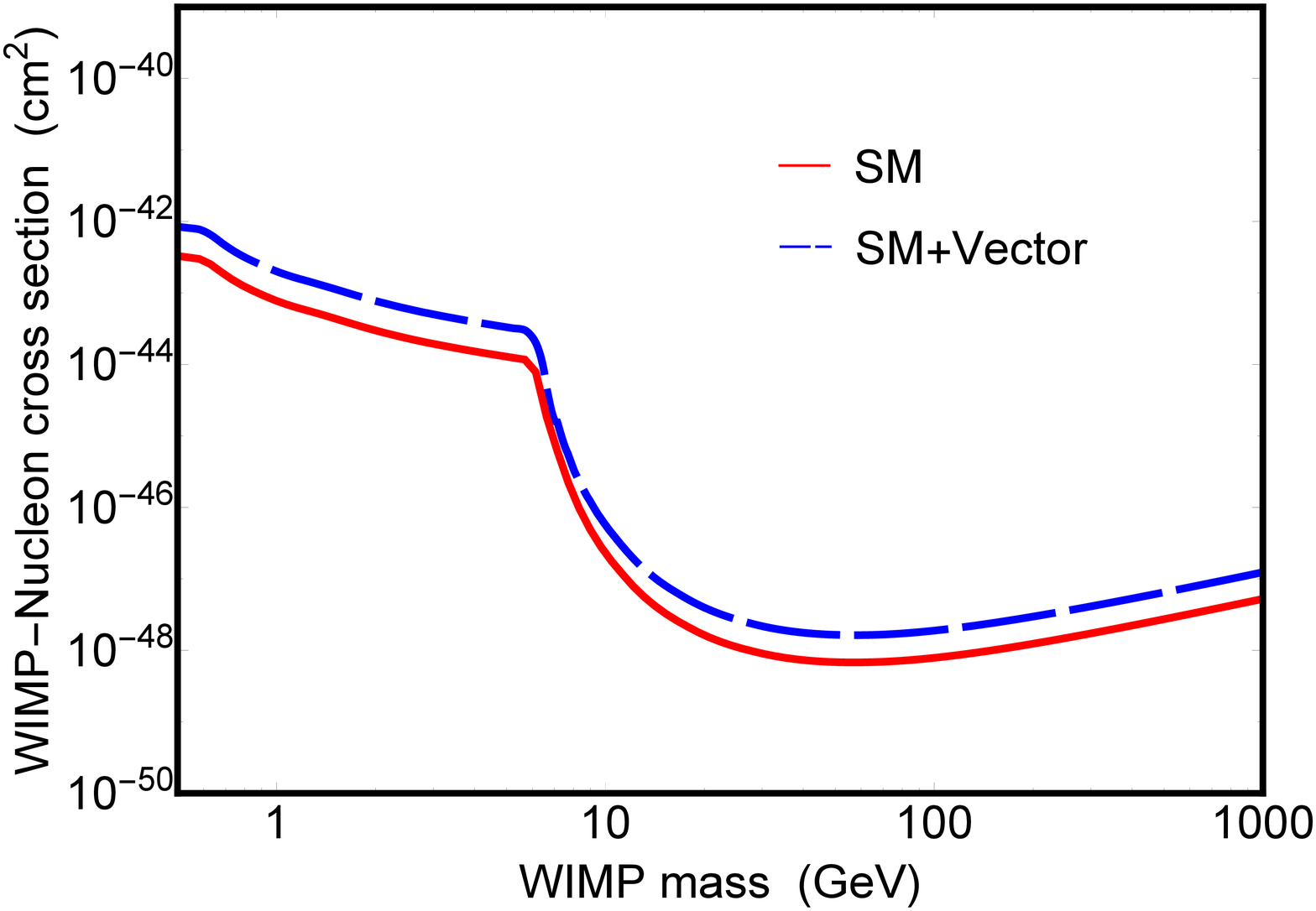}
\hspace{0.5cm}
\includegraphics[width=0.45\textwidth]{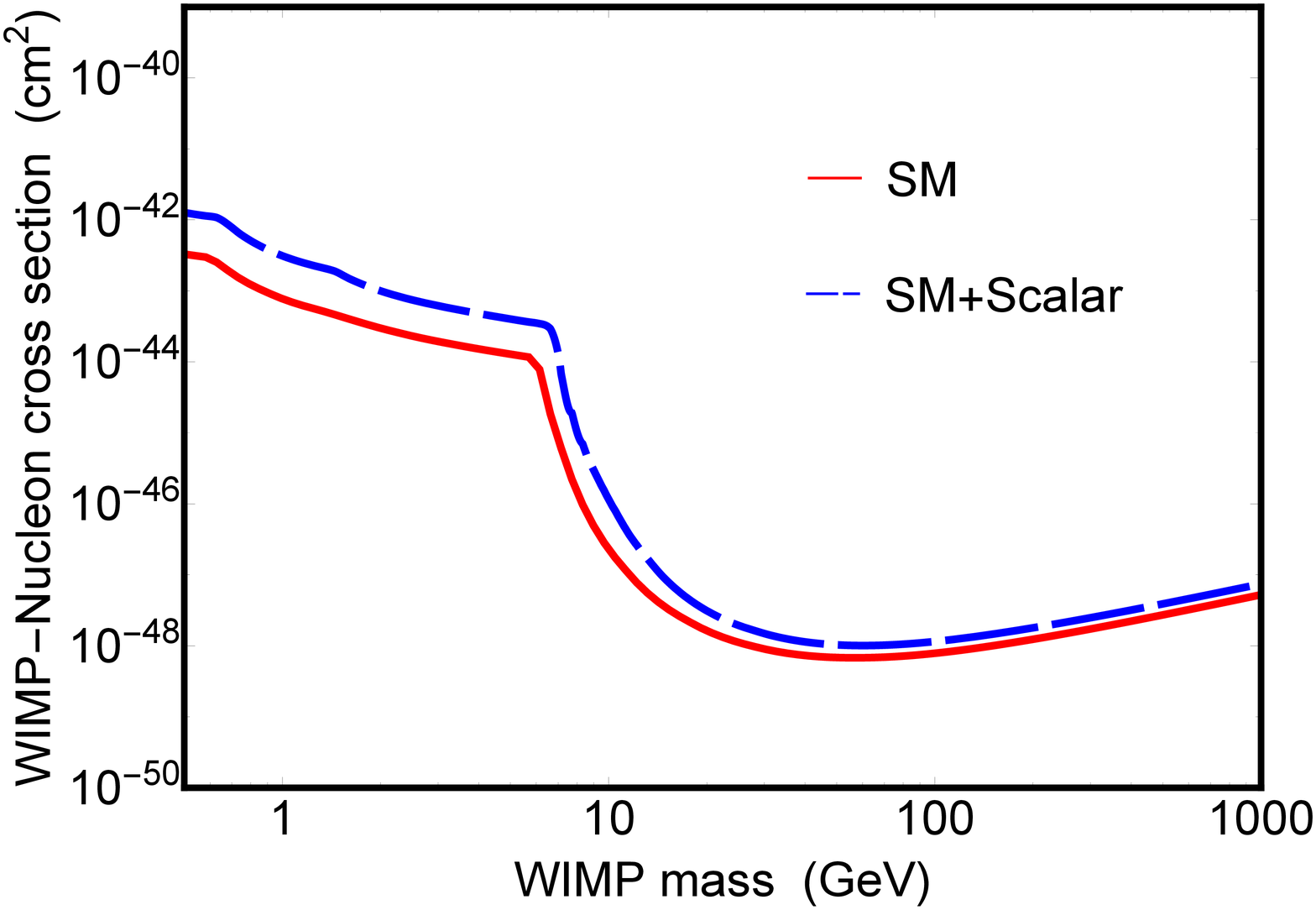}
\caption{First row: event rate as the function of the recoil energy in the Ge72 target in the presence of a single vector current NSI (left-panel) and single scalar current NSI (right-panel). Second row: number of event as the function of energy threshold in the presence of a single vector current NSI (left-panel) and single scalar current NSI (right-panel).  Third row: neutrino floor as the function of dark matter mass  in the presence of a single vector current NSI (left-panel) and single scalar current NSI (right-panel).  } \label{fig:eventratege}
\end{figure}

For completeness, we show in the Fig.~\ref{fig:eventratege}  impacts of the vector-current NSI (left-panel) and scalar current NSI (right-panel) to the neutrino floor in Ge72 target.  In the first, second and third rows we show neutrino event rate as the function of recoil energy, number of $\text{CE}\nu \text{NS}$ event as the function of energy threshold and the neutrino floor as the function of dark matter mass, respectively. Results are similar to  these in the Xe131 case. Since the nuclear response function $W^{\alpha\beta}_{\Sigma', \Sigma^{''}} (q^2)$  for Ge72 are null, contributions of axial-vector-current, pseudo-scalar-current and tensor-current NSIs are zero.

\section{conclusion}

It is well-known that there are limitations on the discovery potential of dark matter in direct detection experiments, the so-called neutrino floor. In this paper, we have examined impactions of non-standard neutrino interactions to the neutrino floor. Our results show that the neutrino floors can be significantly changed by the vector-current NSI and scalar-current NSI, it can be enhanced about ${\cal O}(20\%)$ by the a pesudo-scalar-current NSI,  when considering the current upper bounds  on couplings of NSIs. Contributions of axial-vector-current NSI and tensor-current NSI to the neutrino floor are negligible. Our study shows that one can not exactly determine whether or not it is  a dark matter signal if an event is observed in the future at above the SM neutrino floors but at below our new neutrino floors. In this case, one needs to combine all constraints from neutrino experiments to discriminate the neutrino signal from the dark matter signal. No matter what it is, it will be a signal of new physics but one needs more work to reveal its nature.  One will be pretty sure about the dark matter nature of the signal lying above our new neutrino floors.

\begin{acknowledgments}
This work was supported by the National Natural Science Foundation of China under grant No. 11775025 and the Fundamental Research Funds for the Central Universities under grant No. 2017NT17.
\end{acknowledgments}

\appendix

\section{Nuclear response function}
The nuclear response functions $W_{M,\Sigma',\Sigma''}^{\alpha\beta}$ of Ge72 and Xe131 are shown as following, where $y = (qb/2)^2, b[fm]=\sqrt{41.467/(45A^{-1/3}-25A^{-2/3})}$  with $A$ the isotope of interest. They are taken from the public code ``\textbf{dmformfactor}" given in Ref.~\cite{Anand:2013yka}.

Ge72:
\begin{eqnarray*}
W_{M}^{00}(y)&=&e^{-13.9963y}(103.126-2092.16y+16245.9y^{2}  \nonumber \\
  &-&61392.4y^{3}+119994y^{4}-116348y^{5}+45218y^{6} \nonumber \\
  &-&1304.96y^{7}+9.79078y^{8})\nonumber \\
W_{M}^{01}(y)&=&e^{-13.9963y}(-11.4565+278.503y-2536.5y^{2}+11080.5y^{3}\nonumber \\
&+&24762.7y^{4}+27262.6y^{5}-12055.2y^{6}++518.725y^{7})\nonumber \\
W_{M}^{10}(y)&=&W_{M}^{01}(y)\nonumber \\
W_{M}^{11}(y)&=&e^{-13.9963y}(1.27272-36.0585y+383.661y^{2}-1940.43y^{3}\nonumber \\
&+&4984.81y^{4}-6276.89y^{5}+3186.13y^{6})\nonumber \\
&-&182.418y^{7}+2.77608y^{8}\nonumber
\\
W_{\Sigma'}^{00}(y)&=&W_{\Sigma'}^{01}(y)=W_{\Sigma'}^{10}(y)=W_{\Sigma'}^{11}(y)=0\nonumber \\
W_{\Sigma''}^{00}(y)&=&W_{\Sigma''}^{01}(y)=W_{\Sigma''}^{10}(y)=W_{\Sigma''}^{11}(y)=0\nonumber
\end{eqnarray*}
Xe131:
\begin{eqnarray*}
W_{M}^{00}(y)&=&e^{-16.6234y}(1365.52-45282.6y+592353y^{2}  \nonumber \\
  &+&3.97599\times10^{6}y^{3}+1.5125\times10^{7}y^{4}-3.38604\times10^{7}y^{5}+4.45089\times10^{7}y^{6} \nonumber \\
  &-&3.2834\times10^{7}y^{7}+1.20974\times10^{7}y^{8}-1.68685\times10^{6}y^{9}+76997.6y^{10})\nonumber \\
W_{M}^{01}(y)&=&e^{-16.6234y}(-239.705-9479.45y+154341y^{2}+1.24433\times10^{6}y^{3}\nonumber \\
&+&5.64439\times10^{6}y^{4}+1.4972\times10^{7}y^{5}-2.32082\times10^{7}y^{6}+2.01088\times10^{7}y^{7}\nonumber \\
&-&8.69448\times10^{6}y^{8}+1.43961\times10^{6}y^{9}+76997.6y^{10})\nonumber \\
W_{M}^{10}(y)&=&W_{M}^{01}(y)\nonumber \\
W_{M}^{11}(y)&=&e^{-16.6234y}(42.0782-2027.49y+38307.4y^{2}-368419y^{3}\nonumber \\
&+&1.99059\times10^{6}y^{4}+6.27412\times10^{6}y^{5}+1.15354\times10^{7}y^{6}\nonumber \\
&-&1.1857\times10^{7}y^{7}+6.08587\times10^{6}y^{8}-1.19237\times10^{6}y^{9}+76997.6y^{10})\nonumber \\
W_{\Sigma'}^{00}(y)&=&e^{-16.6234y}(0.0147078-1.1414y+32.683y^{2}\nonumber \\
&-&446.814y^{3}+3401.36y^{4}-14774.3y^{5}+35758.4y^{6}-44223.9y^{7}+21252.4y^{8}\nonumber \\
&+&962.029y^{9}+11.0398y^{10})\nonumber \\
W_{\Sigma'}^{01}(y)&=&e^{-16.6234y}(-0.0139715+1.08922y-31.6241y^{2}\nonumber \\
&+&443.515y^{3}-3460.69y^{4}+15318.3y^{5}-37533.7y^{6}+46711.4y^{7}-22491.7y^{8}\nonumber \\
&-&990.344y^{9}-11.0398y^{10})\nonumber \\
W_{M}^{10}(y)&=&W_{M}^{01}(y)\nonumber \\
W_{\Sigma'}^{11}(y)&=&e^{-16.6234y}(0.0132721-1.0394y+30.5993y^{2}\nonumber \\
&-&439.942y^{3}+3518.97y^{4}-15877.2y^{5}+39392.9y^{6}-49337.7y^{7}+23804.5y^{8}\nonumber \\
&+&1018.66y^{9}+11.0937y^{10})\nonumber \\
W_{\Sigma''}^{00}(y)&=&e^{-16.6234y}(0.00735391+0.199593y\nonumber \\
&-&1.3918y^{2}-21.3512y^{3}+487.152y^{4}-3498.26y^{5}+11846.8y^{6}-19238.3y^{7}\nonumber \\
&+&12306.7y^{8}-116.765y^{9}+0.281202y^{10})\nonumber \\
W_{\Sigma''}^{01}(y)&=&e^{-16.6234y}(-0.00698576-0.195462y\nonumber \\
&+&1.27658y^{2}+21.2236y^{3}-471.84y^{4}+3383.00y^{5}-11490.7y^{6}+18739.6y^{7}\nonumber \\
&-&12055.6y^{8}+115.632y^{9}-0.281202y^{10})\nonumber \\
W_{\Sigma''}^{10}(y)&=&W_{\Sigma^{\prime\prime}}^{01}(y)\nonumber \\
W_{\Sigma''}^{11}(y)&=&e^{-16.6234y}(0.00663605+0.191245y\nonumber \\
&-&1.15856y^{2}-21.2101y^{3}+457.857y^{4}-3274.86y^{5}+11150.6y^{6}\nonumber \\
&-&18258.2y^{7}+11811.3y^{8}-114.498y^{9}+0.2812y^{10})\nonumber
\end{eqnarray*}

\end{document}